\documentclass[a4paper,oneside,english,oneside,reqno]{amsart}
\usepackage[T1]{fontenc}
\usepackage[latin9]{inputenc}
\usepackage{float}
\usepackage{amsthm}
\usepackage{amstext}
\usepackage{graphicx}

\makeatletter
\numberwithin{equation}{section} 
\numberwithin{figure}{section} 
\theoremstyle{plain}

\usepackage[numbers,sort]{natbib}

\@ifundefined{showcaptionsetup}{}{%
 \PassOptionsToPackage{caption=false}{subfig}}
\usepackage{subfig}
\makeatother

\usepackage{babel}

\begin{document}

\title{Prediction by Compression}

\author{Joel Ratsaby}

\address{Department of Electrical and Electronics Engineering, Ariel University
Center, Ariel 40700, ISRAEL}

\email{\texttt{ratsaby@ariel.ac.il} }
\begin{abstract}
It is well known that text compression can be achieved by predicting
the next symbol in the stream of text data based on the history seen
up to the current symbol. The better the prediction the more skewed
the conditional probability distribution of the next symbol and the
shorter the codeword that needs to be assigned to represent this next
symbol. What about the opposite direction ? suppose we have a black
box that can compress text stream. Can it be used to predict the next
symbol in the stream ? We introduce a novel criterion based on the
length of the compressed data and use it to predict the next symbol.
We examine empirically the prediction error rate and its dependency
on some compression parameters.
\end{abstract}

\keywords{Statistical prediction, data compression, binary sequence prediction}

\maketitle

\section{\label{sec:Introduction}Introduction}

Data compression \cite{Sayood2000} is an area in the field of information
theory that studies ways of obtaining efficient representations of
information. Given some general data, for instance, text, picture,
movie, represented in some form of a binary computer file the aim
in data compression is to obtain a new shorter binary description
of this data. This is done by assuming or searching for a structure
in the data that permits removing unimportant repetitions or redundancy
that exists in the original representation of the data. The development
of data compression algorithm can be divided into two phases. The
first phase is referred to as modelling. In this phase we try to remove
any redundancy that exists in the data and describe it in the form
of a model. The second phase is coding. In this phase a description
of the model and a description of how the data differs from the model
are encoded. Producing this description is based on a prediction step.
For instance, in lossless image compression the CALIC algorithm decides
what will be the value of the next pixel in a picture, then encodes
the difference between this predicted value and true value using Huffman
or arithmetic coding. Another example is in text compression: the
PPM algorithm computes the probability distribution of the next symbol
and uses arithmetic coding to encode the value.

In this paper we pose the following question: can a compressor be
used for prediction ? that is, given some compressor can we use it
as a black-box (i.e., not changing its internal workings) to predict
the next symbol in a data sequence of symbols ? The motivation behind
this question is mainly theoretical since practically one can design
a predictor by tackling the problem directly without the need for
data compression. For instance, by modeling the source that produces
the data and doing statistical estimation to learn the model's parameters
and then use the estimator to predict the next symbol given the observed
part of the sequence. In contrast to other works \cite{Choi2003}
that consider compression as a means of doing inference by expanding
on Rissanen's minimum description length principle we take a direct
approach that uses a 'ready-made' compression algorithm as stage in
our inference. Theoretically, the question that we pose is interesting
since prediction by compression, if it is possible, would further
strengthen a basic principle of pattern recognition and statistical
learning theory. The principle is known as Occam's razor (see \cite{Rissanen89,Occam87}):
it says that given several possible explanations for an observed data
one should choose the simplest explanation that is consistent with
the data. 

As far as prediction is concerned we may interpret this principle
as follows: let $\mathcal{A}=\left\{ a_{1},a_{2},\ldots,a_{M}\right\} $
be a finite alphabet consisting of $M$ symbols and denote by $\mathcal{A}^{*}$
the space of all finite sequences of symbols. Given a sequence $x\in\mathcal{A}^{*}$
representing some observed data $x=s_{1}s_{2}\cdots s_{n}$, $s_{i}\in\mathcal{A}$,
$1\le i\leq n$, then according to this principle one should predict
the next symbol $y$ as one which requires the shortest additional
description given $x$. The word 'shortest' is crucial since it is
the basis of Occam's principle. But how should one obtain the shortest
description of a data sequence ?

An obvious approach is to use data compression. Let us denote by $[xy]$
the concatenation of a sequence $x$ of symbols from $\mathcal{A}$
with some symbol $y\in\mathcal{A}$. Denote by $\Xi=\left\{ 0,1\right\} $
and let $\Xi^{*}$ denote the space of all finite binary encoding
sequences. Suppose we have a 'data compression box' represented as
a mapping $\mathcal{C}:\:\mathcal{A}^{*}\rightarrow\Xi^{*}$ from
the space of symbol sequences to their unique binary encodings. Consider
several candidate sequences $[xy_{1}]$, $[xy_{2}]$, $\ldots$, $[xy_{M}]$,
representing all the possible continuations of the data sequence $x$.
Compress each one and measure the length $\ell(\mathcal{C}([xy_{j}]))$,
$1\leq j\leq M$, of their corresponding binary encodings.  Then the
idea is to predict the next symbol to be $y_{j^{*}}$ where $j^{*}=\text{argmin}_{1\leq j\leq M}\ell(\mathcal{C}([xy_{j}]))$. 

This is our theoretical interpretation of Occam's principle for the
prediction paradigm. The current paper aims to check if this interpretation
works in reality.

\section{Setup}

Our setup consists of a source capable of producing random binary
sequences. The source is based on a Markov chain with states $s\in\left\{ 0,1,\ldots,N-1\right\} $
which we represent as the set of binary vectors $\left\{ \left(0,\ldots,0\right),\left(0,0,\ldots,0,1\right),\ldots,\left(1,1,\ldots,1\right)\right\} $,
respectively, using $\log_{2}N$ bits. This number of bits is referred
to as the \emph{order} of the source which we denote by $\rho$. We
alternatively use the decimal or binary representations of states.
Given that the source is at a state $b=\left(b(0),\ldots,b(k-1)\right)\in\left\{ 0,1\right\} ^{\rho}$
then it can change into one of only two states $\left(b(1),b(2),b(3),\ldots,b(k-1),0\right)$
or $\left(b(1),b(2),b(3),\ldots,b(k-1),1\right)$. We denote the transitions
into these states as the $0$-transition and $1$-transition, respectively.
Associated with state $i$ there is a conditional probability distribution
$[P(1|i),P(0|i)]$ which gives the probability of making a $0$-transition
or a $1$-transition. The source produces a finite binary sequence
$x$ of length $n$. It will be convenient to refer to the $i^{th}$
bit of $x$ as the bit value $x(i)$ at discrete time $i$.

The prediction problem that we consider is incremental. We start by
having the first $m$ bits of $x$ as observables and predict the
$(m+1)^{th}$ bit. After seeing the true value $x(m+1)$, we add it
to the observables and predict the $(m+2)^{th}$ bit. This iterative
process is repeated until we finish predicting the $n^{th}$ bit.
Let us denote by $D(i)$ the binary value decided by the predictor
at time $i$, where $m+1\leq i\leq n$. In predicting the value of
a bit $x(i)$, the predictor makes an error if its decision $D(i)$
at time $i$ differs from $x(i)$. As a measure of goodness of the
predictor, we estimate the probability that it makes an error in the
future as the number of errors that it makes on the $n-m$ bit predictions
divided by $n-m$. We denote this estimate as $P(err)$.

As discussed in section \ref{sec:Introduction} we are considering
a predictor that bases its decision on a compressor. For compression
we use the PPMd algorithm \cite{Sayood2000,PPM}. The PPMd algorithm
is based on a predictive model that estimates the probability of the
next symbol and encodes it by arithmetic coding. The specific implementation
that we use (called 7zip) permits to set the maximal context size
$d$ of the PPM algorithm up to a value of $32$ with a default value
of $6$ (this is the number of symbols that it uses as the given observables
to estimate the probability of the next yet unseen symbol). We will
use $d$ as one of the parameters that controls the predictor. Another
parameter is the history size $h$. We define a window of length $h$
bits. At prediction time $i$, $m+1\leq i\leq n$, instead of using
all previous $i-1$ bits as observables we let the predictor see only
the last $h$ bits of $x$ at time instants $i-h$, $i-h+1$, $\ldots$,
$i-1$. We denote this by $x_{h}(i)$ and refer to this subsequence
as the 'history' for prediction time $i$. In our tests, we consider
different values for the parameters $d$ and $h$. Let $\mathcal{C}(x)$
be the binary string obtained from compressing the binary sequence
$x$ and $\ell(x)$ the length of $x$. Write $\lambda(x)$ for $\ell\left(\mathcal{C}\left(x\right)\right)$.

\subsection{\label{sub:Initial-investigation}Initial investigation}

We chose a source with $N=4$ and conditional probability distribution
$\left[P(1|i),P(0|i)\right]=\left[0.3,0.7\right]$ for each of the
four states $0\leq i\leq3$. Here the Bayes optimal decision is $0$
at every state and the Bayes error is $0.3$. For the compressor we
used PPMd without controlling the $d$ parameter (the maximal possible
context size is left to be the default size of $6$). Our first attempt
at a compression-based predictor was to try to predict not just the
next single bit of $x$ but a whole future subsequence $y$ (of length
$\ell(y)=l$) of $x$. With each prediction time $i$ we associated
two counters $\alpha_{i},\beta_{i}$ which keep the number of times
that we covered bit $i$ and the number of times we predicted the
value $1$ for this time instant $i$, respectively. We divided the
prediction process into two phases. In the first phase initialize
all counts to zero then at each prediction time we considered $2^{l}$
prediction candidates and chose the one with the best score value
(score is specified below). Once a subsequence $y_{i}=\left(y_{i}(0),\ldots y_{i}(l-1)\right)$
was decided at time $i$ we incremented by one the counts $\alpha_{t}$
of all covered time instants $i\leq t\leq i+l-1$ and incremented
by one the counts $\beta_{t}$ corresponding to those time instants
in the range $i\leq t\leq i+l-1$ whose value $y_{i}(t-i)=1$. We
left the counts of the remaining time instants unchanged. This was
repeated for all prediction times $m+1\leq i\leq n-l$. Then the second
phase started. For all time $m+1\leq i\leq n$, decide the value $D(i)=1$
if $\beta_{i}\geq\alpha_{i}/2$ else decide $D(i)=0$.

We let the candidate length $l$ in bits be a parameter and together
with the parameter $h$ (describe above) we repeated the prediction
over many runs where for each run we set a different parameter value
$\left(h,l\right)$, $100\leq h\leq2000$, in increments of $100$
and $1\leq l\leq8$. 

We used the following function to score a prediction candidate subsequence
$y$ given the history $x_{h}(i)$ for all times $i$: \begin{equation}
S(x_{h},y)=\lambda([x_{h}y])\label{eq:s1}\end{equation}
where $\left[x_{h}y\right]$ denotes the concatenation of the strings
$x_{h}$ and $y$. This score represents the description length necessary
for describing $y$ and the observed $x_{h}$ . With Occam's principle
the intuition says that the lower $S(x_{h},y)$ the better the candidate
$y$. However this scheme led to unsatisfactory results as the prediction
error was significantly higher than Bayes error for all runs. Another
score function that we tried is

\begin{equation}
S(x_{h}:y)=\lambda(x_{h})-\left(\lambda([yx_{h}])-\lambda(y)\right).\label{eq:s2}\end{equation}

We may regard $\lambda([yx_{h}])-\lambda(y)\equiv\lambda(x_{h}|y)$
as the extra description length needed for describing $x_{h}$ given
the knowledge of $y$ (extra meaning beyond what is needed to describe
just $y$). Note $\lambda(x_{h}|y)$ resembles Kolmogorov's combinatorial
notion of conditional entropy \cite{Kolmogorov65} (see also \cite{RATSABY_DBLP:conf/sofsem/Ratsaby07,Ratsaby_IW}).
$S(x_{h}:y)$ resembles Kolmogorov's combinatorial information measure.
Indeed $S(x_{h}:y)$ measures the information that $y$ provides about
$x_{h}$since it shows the reduction in description length of $x_{h}$
once $y$ is used as an input for describing $x_{h}$. However still,
this did not improve the previous results.

At this point we decided to change the source. We increased the number
of states to $8$ still with only two possible transitions outgoing
from each state with probability $0.3$ and $0.7$. Thus the Bayes
error is $0.3$. This made the source more complex, however, unlike
before, here we used $8$ bit combinations chosen at random (and not
$3$ bits) to represent each of the states and the sequence $x$ was
now generated based on the state combinations in a non-overlapping
manner. This introduced redundancy in the sequence representation.
For instance, suppose the state combinations are the strings $00000001$,
$00000010$, $\ldots$, $10000000$ and suppose that the transitions
are obtained by shifting in a circular manner the $1$ bit to the
left or right by one position. Then an example of a sequence $x$
is: $00000001\:$$00000010\:$$00000001\:$$100000000\ldots$. Such
a scheme led to some improvement but the minimal error obtained was
still above $0.4$ hence larger than Bayes error.

\subsection{Injecting redundancy by transition-encoding}

At this point we had some more insight, mainly, that adding redundancy
to the source somewhat improves prediction by compression. But in
general one cannot change the source of the data. So we chose to leave
the source as generating the data sequence $x$ based on single bit
transitions, i.e., $0$-transition and $1$-transition, as described
in the first paragraph on section \ref{sub:Initial-investigation}.
The idea at this stage was to extract the information from $x$ and
form a new sequence based on $k$-bit words of $x$ where the value
of $k$ is fixed in advance in the range $2\leq k\leq7$. We call
this process \emph{peeling}. Let us describe it through an example:
suppose $x=1010100$ and suppose $k=4$ then the extracted words are
(in sequence): $1010$, $0101$, $1010$, $0100$. We then complete
each such \emph{$k$-codeword} into an $8$-bit word (a byte) by inserting
$8-k$ zeros in the front. For this example it yields $00001010$,
$00000101$, $00001010$, $00000100$. We then create a random injective
map from the set $0^{8-k}\times\left\{ 0,1\right\} ^{k}$ to $128$
characters whose UTF-8 representation fall in the range U+$0000$
to $U+007F$ (we write $0^{l}$ to denote a string of $l$ zeros).
Note that for this range of UTF-8 each character is represented by
a single byte. This is important since when we compare candidates
for prediction we base it on the compressed length of the representation
(in this the representation is UTF-8 encoded characters). Having chosen
to limit to single-byte UTF-8 representation ensures that any two
candidate sequences made up of the same number of characters will
be encoded with the same length encoding sequences.

We map $x$ into a sequence of characters, for instance, the above
sequence could be mapped into $ADAG$. Each symbol represents a state
(we call them \emph{perceived states }since they are from the perspective
of the predictor). The number of such states is $2^{k}$ and it need
not necessarily equal the number $N$ of source's states since $k$
may differ from $\rho$ (the order of the source). We then convert
this character sequence into a sequence $\chi$ specifying the pairwise
transitions (except the last entry which is half full) and repeating
each transition $r$ times (we term it the \emph{repeat factor}).
We denote this as \emph{transition-encoding} scheme. For this example,
suppose $r=3$ then it takes the following form

\begin{eqnarray}
\lefteqn{\chi_{h}^{(r)}=A\Rightarrow D,A\Rightarrow D,A\Rightarrow D,}\nonumber \\
 & D\Rightarrow A,D\Rightarrow A,D\Rightarrow A,A\Rightarrow G,A\Rightarrow G,A\Rightarrow G,G\Rightarrow\label{eq:xhi}\end{eqnarray}
where the symbol $\Rightarrow$ has a UTF-8 code $0\mathsf{x}BB$.
Note, the last transition (the trailer) is half full, in this example
it is $G\Rightarrow$, and consists of the state (in this case $G$)
on the right of the last transition ($A\Rightarrow G$). Thus $\chi^{(r)}$
is a sequence of bytes that encode perceived-state characters, a comma
or the arrow symbol. Clearly, $\chi^{(r)}$ has more redundancy than
$x$ since its length is $O(kr)$ times longer than $x$ yet it contains
the same amount of information. Obviously if the source's order $\rho$
is known to the predictor then a good choice is to let $k=\rho$.
If it is not known, this method can still be used by scanning over
all values $2\leq k\leq7$ while testing the probability of error
(on some training sequence $x$). This way a value for $k$ will be
selected when the minimal probability of error is obtained over this
range.

Can the score function defined in (\ref{eq:s2}) be used here ? Instead
of $x_{h}$ we have $\chi_{h}$ so we need to represent prediction
candidates $y$ in the UTF-8 symbol sequence format, denoted as $\xi$,
of the from $A\Rightarrow B$ . Only after converting $y$ into $\xi$
we may attempt to concatenate them to form an UTF-8 symbol sequence
$\xi\chi$. In this new representation we are coding transitions between
states rather than absolute states. Prediction candidates $\xi$ should
represent a transition starting from the state that corresponds to
the symbol on the right of the last pairwise transition of $\xi$.
For instance in the above example, this last pairwise transition is
$A\Rightarrow G$ hence we should only consider the two candidates
$\xi$ that code transitions of the from $G\Rightarrow\cdot$ where
$\cdot$ is a symbol whose $k$-bit codeword starts with the first
$k-1$ bits of the codeword of $G$ and ends either with a $0$ or
$1$ (denote them as the $\xi_{0}$ and $\xi_{1}$ candidates). But
to form the concatenation $\xi\chi_{h}$ means that the right symbol
of these two candidates must be the left symbol of the first transition
of $\chi_{h}$ (in the above example this is the symbol $A$). Hence
there may be a conflict constraints in determining the two candidates
since as in this example the symbol $A$ may not be a symbol whose
codeword agrees on the first $k-1$ bits with the symbol $G$. Hence
the score (\ref{eq:s2}) is not suitable. 

This led us to consider the following new scoring function,

\begin{equation}
S(\xi:\chi_{h})=\lambda(\xi)-\left(\lambda([\chi_{h}\xi])-\lambda(\chi_{h})\right)\label{eq:syx}\end{equation}
Note that we may regard $\lambda([\chi_{h}\xi])-\lambda(\chi_{h})\equiv\lambda(\xi|\chi_{h})$
as the extra description length needed for describing $\xi$ given
the knowledge of $\chi_{h}$. In a similar way as we interpreted $S(x_{h}:y)$
above, here $S(\xi:\chi_{h})$ is what Kolmogorov \cite{Kolmogorov65}
calls the information that $\chi_{h}$ provides about $\xi$. Note
that to evaluate (\ref{eq:syx}) we just need to convert $y$ into
$\xi$ and $x_{h}$ into $\chi_{h}$ (using the above process). Unlike
the string $\xi\chi_{h}$ considered above, now the concatenated string
$\chi_{h}\xi$ is well-defined since we just look at the UTF-8 symbol
of the trailer of $\chi_{h}$ (in the above example it is $G$) and
only consider the two candidates $\xi$ that encode transitions of
the form $G\Rightarrow\cdot$ where $\cdot$ is a symbol whose $k$-bit
word starts with the last $k-1$ bits of the codeword of $G$ and
ends either with a $0$ or $1$. In the above example, suppose that
corresponding to the $0$ and $1$ candidates the two candidate symbols
are $V$ and $W$ then we will form the candidate sequences\begin{eqnarray*}
\xi_{0} & = & G\Rightarrow V\\
\xi_{1} & = & G\Rightarrow W\end{eqnarray*}

Note that the candidates sequences depend on the trailer of $\chi_{h}$
(in this example it is the symbol $G$). When we form the concatenated
sequence $\chi_{h}\xi$ we first remove the trailer of $\chi_{h}$
and then attach $\xi$ form the right. Hence in the above example
we get \begin{eqnarray*}
\lefteqn{\chi_{h}^{(r)}\xi_{0}=A\Rightarrow D,A\Rightarrow D,A\Rightarrow D,}\\
 & D\Rightarrow A,D\Rightarrow A,D\Rightarrow A,A\Rightarrow G,A\Rightarrow G,A\Rightarrow G,G\Rightarrow V\end{eqnarray*}
Using this representation scheme we did several runs with a source
of order $\rho=3$ and a predictor with peeling size $k=3$ . We obtained
encouraging results with the prediction error being smaller than in
the previous trials but staying around $0.4$ hence not close enough
to the Bayes error even when the history size $h$ is large. 

We then observed the second term of the score function (\ref{eq:syx})
for several sample runs and saw that most of the times both candidates
$\xi_{0}$, $\xi_{1}$ have the same value for compressed length $\lambda([\chi_{h}\xi_{0}])=\lambda([\chi_{h}\xi_{1}])$
and $\lambda(\xi_{0})=\lambda(\xi_{1})$. This makes them be indistinguishable
which leads to random prediction (since we we predict a $1$ with
probability $0.5$ when the scores are the same). We realized that
this is a result of the compression algorithm imperfectness. It has
some constant minimal file size that it never goes below no matter
how short the sequence to be compressed is. Hence the next idea was
to consider extending the length of each candidate $\xi$ by repeating
it $qr$ times (where $r$ is the repeat factor for $\chi_{h})$.
For instance, in the above example (\ref{eq:xhi}) where $r=3$ using
$\xi_{0}=G\Rightarrow V$, then if $q=2$ its extension is \begin{eqnarray*}
\xi_{0}^{(qr)} & = & G\Rightarrow V,G\Rightarrow V,G\Rightarrow V,G\Rightarrow V,G\Rightarrow V,G\Rightarrow V\end{eqnarray*}
Now, the lengths of the compressed version of the two strings $\lambda(\chi_{h}^{(r)}\xi_{0}^{(qr)})$,
$\lambda(\chi_{h}^{(r)}\xi_{1}^{(qr)})$ become distinguishable (in
our experiments we used $r=20$ and $q=100$). Doing that led to significant
improvement and we can now see the prediction error decreasing towards
the Bayes error with increasing history size $h$ and increasing PPM
order $d$. Note that we are still treating the case where the peeling
size equals the source order (which is less-realistic since the predictor
may not know the source's order).

\subsection{The predictor}

Let us summarize the features of the final version of the predictor:
\begin{itemize}
\item \emph{Aim}: Given history $x_{h}$ at bit time $i$ , we want to predict
bit $i$, $m+1\leq i\leq n$
\item Peel $k$-codewords from $x_{h}$ and map them into UTF-8 symbols
representing predictor's perceived states
\item Represent the transitions between these states in the form $A\Rightarrow B$
and repeat each $r$ times to form the sequence $\chi_{h}^{(r)}$.
\item There are two prediction candidates, $\xi_{0}$, $\xi_{1}$ corresponding
to the two possible binary predictions of the next bit in $x_{h}$.
We use the same representation of state transition to represent the
two candidates. We use a repeat factor of $qr$ to obtain $\xi_{0}^{(qr)}$,
$\xi_{1}^{(qr)}$.
\item Use the scoring function $S(\xi:\chi_{h})$ to score each of the two
candidates. If both have the same value then predict $1$ with probability
$0.5$.
\item Otherwise, predict the next bit $i=\text{argmax }_{j\in\left\{ 0,1\right\} }S(\xi_{j}^{(qr)}:\chi_{h}^{(r)})$
\end{itemize}
The next series of trials were more realistic as we allowed the predictor's
peeling size $k$ to be different from the source's order. Let us
review the results.

\section{Results}

\begin{figure}[H]
\begin{centering}
\subfloat[\label{fig:1a}$P(err)$ versus PPM order $d$]{\noindent \begin{centering}
\includegraphics[clip,scale=0.7]{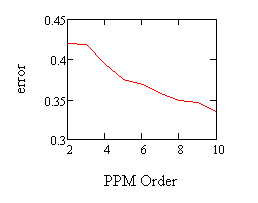}
\par\end{centering}

}
\par\end{centering}

\begin{centering}
\subfloat[\label{fig:1b} $P(err)$ as function of history size $h$, PPM order
$d$]{\begin{centering}
\includegraphics[clip,scale=0.5]{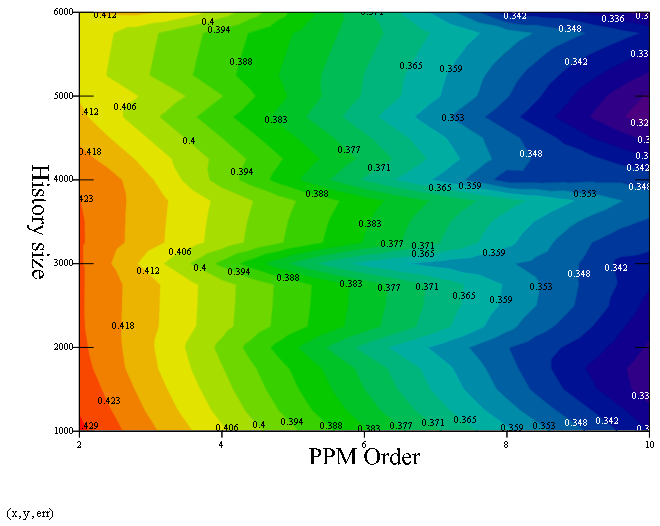}
\par\end{centering}

}
\par\end{centering}

\caption{\label{fig:Source-model-has}Source model has $8$ states ($\rho=3$
bits), predictor's peeling size is $4$. Initially prediction is bad
but as PPM order increases we see steady decrease in error. The predictor
is rich enough as for each $3$-bit source state it has two perceived
states (each $4$ bits long). Hence with sufficiently large order
it is able to reach low error rates.}

\end{figure}

In the first test we considered a source of order $\rho=3$ and peeling
size $k=4$. In this case the number of perceived states $2^{k}=16$
is greater than the number of source's states $N=8$. To see the effect
of the compressor's PPM order $d$ on the predictor's error $P(err)$
we performed several trial runs changing the history size $h$ and
PPM order $d$ and measuring the prediction error $P(err)$. Figure
\ref{fig:1a} shows how $P(err)$ decreases with respect to $d$.
We plot the error for each value of $d$ averaged out over $5$ runs
of the same history size $h$ and where $h=1000$, $2000$ , $\ldots$,
$6000$. At this rate of decrease the prediction error would reach
the Bayes error of $0.3$ at about $d=14$. Note that the error starts
decreasing from its maximal value only at $d=3$. At this value the
PPM compression captures the full trailer of $\chi_{h}$ . Its context
conditional probability distribution for the next character in the
sequence contains full information about the perceived state (in the
above example this is the state $G$). This is a necessary condition
for PPM to estimate the transition probability correctly. Figures
\ref{fig:1a} shows that $P(err)$ decreases steadily as we increase
the PPM order $d$. Figure \ref{fig:1b} shows that there is hardly
any effect on the error as the history size $h$ increases.

\begin{figure}[H]
\begin{centering}
\subfloat[\label{fig:2a}$P(err)$ versus PPM order $d$]{\begin{centering}
\includegraphics[scale=0.7]{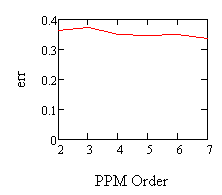}
\par\end{centering}

}
\par\end{centering}

\begin{centering}
\subfloat[\label{fig:2a-1}$P(err)$ as function of history size $h$, PPM order
$d$]{\begin{centering}
\includegraphics[scale=0.5]{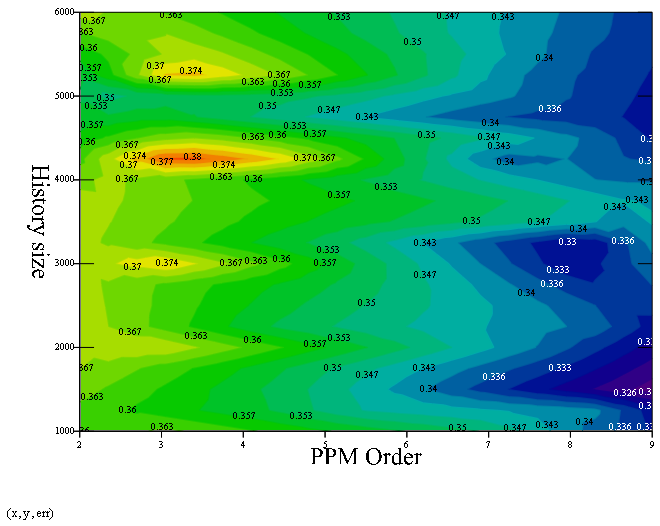}
\par\end{centering}

}
\par\end{centering}

\caption{\label{fig:2}Source model has $16$ states ($\rho=4$), peeling size
is $4$. The problem is easy as can be seen from the low error (close
to Bayes error of $0.3$) obtained already at the starting PPM order
value of $2$.}

\end{figure}

What happens when $\rho=k$ ? We tested a predictor of peeling size
$k=4$ on a sequence produced by a source of order $\rho=4$. Figures
\ref{fig:2a} and \ref{fig:2a-1} show that already from $d=2$ we
see an error close to the Bayes error.

\begin{figure}[H]
\begin{centering}
\subfloat[\label{fig:3a}]{\begin{centering}
\includegraphics[scale=0.7]{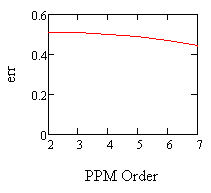}
\par\end{centering}

}
\par\end{centering}

\begin{centering}
\subfloat[\label{fig:3b}]{\begin{centering}
\includegraphics[scale=0.5]{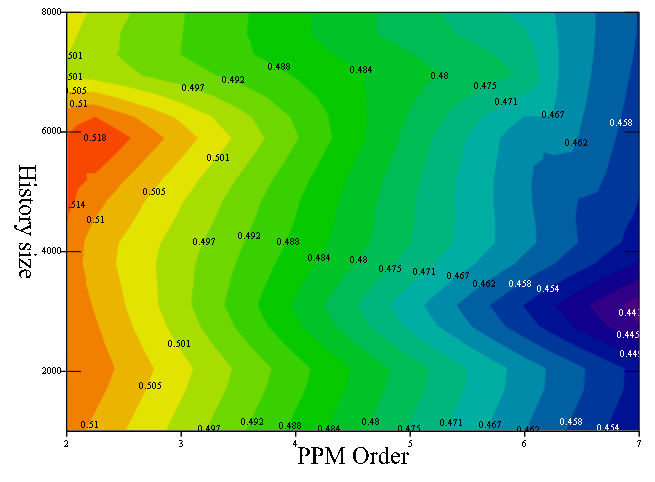}
\par\end{centering}

}
\par\end{centering}

\caption{\label{fig:The-critical-point-1-1}Source model has $32$ states ($\rho=5$
), peel size is $4$. It takes a large value of PPM order $d$ before
we start seeing a prediction error which is slightly lower than pure
guessing (i.e., $0.5$ error rate).}

\end{figure}

Next, we tested a predictor of peeling size $k=4$ on a sequence produced
by a source of order $\rho=5$. Figures \ref{fig:3a} and \ref{fig:3b}
show that the predictor is inaccurate since even when the order $d$
is high we still see an error which is only slightly better than pure
guessing.

\begin{figure}
\subfloat[\label{fig:4a}]{\begin{centering}
\includegraphics[scale=0.7]{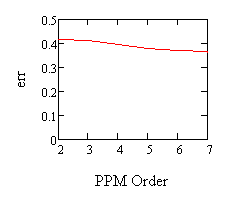}
\par\end{centering}

}

\subfloat[\label{fig:4b}]{\begin{centering}
\includegraphics[scale=0.5]{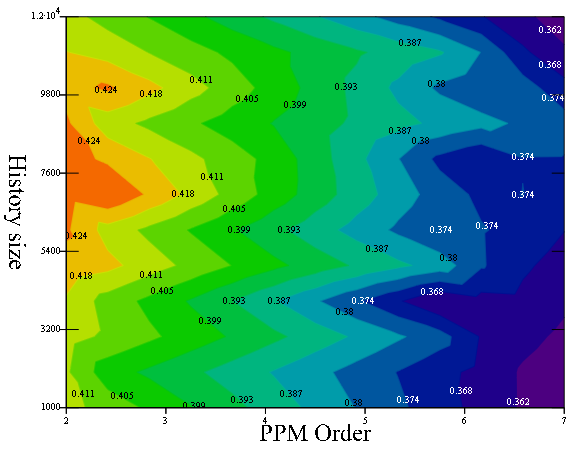}
\par\end{centering}

}

\caption{\label{fig:The-critical-point-1-1-1}Source model has $32$ states
(each $5$ bits), peel size is $6$. Initially prediction is bad but
as PPM order increases we see steady decrease in error. The predictor
is rich enough as for each source state it has two combinations (each
$5$ bits long). Hence with sufficiently large order he is able to
reach low error rates but the rate with respect to PPM order is not
as fast as in Figure \ref{fig:Source-model-has-1}.}

\end{figure}

Next, we tested a predictor of peeling size $k=6$ on a sequence produced
by a source of order $\rho=5$. We let the history size reach up to
$12000$. Figures \ref{fig:4a} and \ref{fig:4b} show that the predictor's
error decreases with increasing PPM order albeit at a slower rate
than the case of $k=4$ and $\rho=3$ (Figures \ref{fig:1a},\ref{fig:1b}).
This suggests that the error rate not only depends on the difference
$|k-\rho|$ (which in both case equals $1$) but also on $\rho$ itself.

\begin{figure}[H]
\begin{centering}
\includegraphics[scale=0.7]{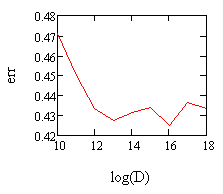}
\par\end{centering}

\caption{\label{fig:Source-model-has-1}Source model has $16$ states ($\rho=4$),
peeling size is $4$. The predictor is based on LZMA compression,
a dictionary-based algorithm. The prediction error is plotted with
respect to $\log_{2}(D)$, where $D$ is size of the dictionary}

\end{figure}

Finally, we changed the compressor's algorithm from PPM to LZMA which
is a recently developed algorithm that incorporates the Lempel-Ziv
dictionary-based compressor with a new predictive component. We tested
the predictor with LZMA on the problem of $\rho=4$ and $k=4$ and
instead of PPM order we used the dictionary size as a parameter. As
Figure \ref{fig:Source-model-has-1} shows, the error initially decreases
but remains above the $0.4$ level even as we increase the dictionary
size at an exponential rate. On this same problem the PPM predictor
performed very well (Figure \ref{fig:2}). We are not sure why the
LZMA underperforms compared to the PPM approach. Perhaps doing prediction
based on dictionary-based compression requires a different transition-encoding
scheme. This requires further investigation.

\section{Conclusions}

In this paper we raise a new question: can a compression algorithm
be used as a 'black-box' to obtain prediction ? We introduced a new
scoring function $S(\xi:\chi_{h})$ which measures the amount of information
that the history $\chi_{h}$ gives about a candidate future prediction
$\xi$. We showed through a series of empirical results that this
scoring function leads to successful prediction by a predictor whose
decision for the next bit is obtained by maximizing the score. Prediction-by-compression
works and agrees with the intuition that says to choose the candidate
future prediction $\xi$ for which the history gives a maximal amount
of information.

\bibliographystyle{plainnat}

\end{document}